\documentclass[fleqn,twoside]{article}
\usepackage{espcrc2}

\usepackage{graphicx}
\usepackage[figuresright]{rotating}

\title{Observations of HI 21cm absorption by the neutral IGM
during the epoch of re-ionization with the Square Kilometer Array}

\author{C.L. Carilli\address{NRAO, Socorro, NM, USA,
ccarilli@nrao.edu}\thanks{The National Radio Astronomy Observatory
(NRAO) is operated by Associated Universities, Inc. under a
cooperative agreement with the National Science Foundation.}, 
N. Gnedin\address{CASA, University of Colorado, Boulder, CO, USA},
S. Furlanetto\address{California Institute of Technology, Pasadena, CA,
USA},
F. Owen\address{NRAO, Socorro, NM, USA}}

\begin{document}

\begin{abstract}

We explore the possibility of detecting HI 21cm absorption by the
neutral intergalactic medium (IGM) toward very high redshift radio
sources, and by gas associated with the first collapsed structures,
using the Square Kilometer Array at low frequency (100 to 200 MHz).
The epoch considered is between the time when the first ionizing
sources form and when the bulk of the neutral IGM becomes ionized.
Expected IGM absorption signal includes $\sim 1\%$ absorption by the
mean neutral IGM (the radio 'Gunn-Peterson' effect'), plus deeper,
narrow lines ($\ge 5\%$, a few km s$^{-1}$) arising in mild density
inhomogeneities with typical values of cosmic overdensity $\delta\sim
10$, precisely the structures that at later times give rise to the
Ly$\alpha$ forest (the `21cm forest').  Absorption can also arise in
gas associated with collapsed structures ($\delta\ge 100$), including
'minihalos' ($\le 10^7$ M$_\odot$) and protodisks ($\ge 10^8$
M$_\odot$). We consider SKA sensitivity limits and the
evolution of radio source populations, and conclude that it is
reasonable to hypothesize the existence of an adequate number of
high-$z$ radio sources against which such absorption studies could be
performed, provided that reionization occurs at $z < 10$.  
Lastly, we discuss the possibility of `line confusion' due
to radio recombination lines arising in the ionized IGM.  
Overall, SKA absorption studies should provide a
fundamental probe of the thermal state of the neutral IGM during the
epoch of reionization, as well as critical insight into the process and
sources of reionization.

\end{abstract}
\maketitle

\section{Introduction}

The epoch of reionization (EoR) presents a key benchmark in the study of
cosmic structure formation, signaling the time when the first luminous
structures form in the cosmos.  Furlanetto et al. (this volume) 
introduce the physical
processes and questions involved in reionization, the current
observational constraints on reionization (Gunn-Peterson absorption
troughs and large scale polarization of the CMB), and the unique
capability of the SKA to study the neutral intergalactic medium (IGM)
via HI 21cm emission observations (or absorption against the CMB).  HI
21cm emission observations probe large scale structure in the IGM ($>
10^{13}$ M$_\odot$), including structure in density, excitation
temperature, and fractional ionization.

In this chapter we describe observations aimed at studying HI 21cm
absorption toward the first radio sources within the EoR.  The neutral
IGM is opaque at rest wavelengths shorter than Ly$\alpha$ such that
study of objects, and the IGM, during this age will be limited to
observations at wavelengths longer than about 1$\mu$m. However, the
weakness of the magnetic hyperfine transition makes the IGM
translucent to HI 21cm absorption.  Column density sensitivity for
absorption studies is set only by the surface brightness of the
background source, thereby allowing absorption studies to probe to
orders-of-magnitude lower 'masses' than can be detected in 21cm
emission.

\section{21cm absorption by the neutral IGM}
\subsection{Simulations of the IGM during reionization}

Carilli et al. \cite{carilli02} present a detailed study of HI 21cm
absorption by the neutral IGM during reionization. We review their
results herein. Their analysis relies on the simulations of Gnedin
\cite{gnedin00}.  These simulations include the three main
physical ingredients required to model neutral hydrogen absorption in
the redshifted 21 cm line: inhomogeneous small-scale structure of the
universe, radiative transfer, and accurate treatment of the level
populations in atomic hydrogen.  The simulations are normalized in
such a way as to reproduce both the observed star formation rate at
$z\sim4$ and the observed evolution of the mean transmitted flux in
the spectra of two $z\sim6$ quasars.  We can therefore expect that
these simulations are at least representative of the physical
processes involved, with the caveat that some of the details, such as
the exact redshift of reionization, or assumptions about radiative
transfer, remain uncertain. In particular, these simulations have not
been adjusted for possible early reionization, as suggested by the
recent WMAP results \cite{kogut03}.

A fundamental parameter in predicting the HI 21cm absorption
characteristics of the IGM is the behavior of the excitation
temperature of the HI (the `spin temperature', $T_S$).  Tozzi et
al. \cite{tozzi00} showed that during the EoR the excitation temperature is
likely to be in excess of the CMB temperature due to the standard
Wouthuysen-Field effect, i.e.\ resonant scattering of the ambient
Ly$\alpha$ photons emitted by the first ionizing sources (at higher
densities collisions with electrons and neutral atoms also play a
role).  As the structure develops, the kinetic temperature $T_K$
increases both due to shock heating of the gas in high density regions
and X-ray heating in low density regions. However, as the
Ly$\alpha$ excitation rate $P_\alpha$ does not increase very fast with
redshift, the spin temperature:
\begin{equation}\label{Ts}
T_S = T_{\rm CMB}{P_\alpha+P_{\rm th}\over P_{\rm th}+P_\alpha 
T_{\rm CMB}/T_K},
\end{equation}
(here $P_{\rm th}=7.6\times10^{-13}{\rm s}^{-1}(1+z)$ is the so-called
``thermalization `` rate \cite{madau97,tozzi00}), 
increases at a slower rate, because the spin temperature becomes
independent of the kinetic temperature when the latter gets very
large. The 21 cm optical depth therefore depends on the evolution of the
kinetic temperature of the IGM, which in turn depends on processes
such as shock heating, the ionization history, and the X-ray
background.  The particular simulation described here is thus one
example of the possible 21 cm forest evolution.

Figure 1 shows physical quantities along a representative line of
sight through the simulation box at three different redshifts ($z = 8,
10, 12$), with redshift indicated by the line color.
The top panel shows the HI 21cm transmissivity of the neutral IGM at
high velocity resolution (0.5 kHz $= 1$ km s$^{-1}$ at 150 MHz).  The
abscissa for this panel is velocity. The middle panel shows the
kinetic and spin temperatures of the gas.  The bottom panel shows the
neutral hydrogen density structure. The abscissa in these two cases is
the corresponding comoving physical scale.

\begin{figure}
\includegraphics[width=3in]{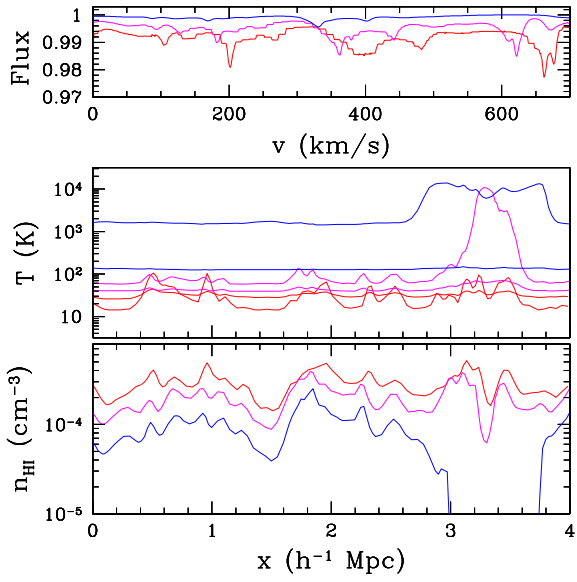}
\caption{ The upper panel shows the transmitted
radio flux density over a relatively narrow velocity range (700 km
s$^{-1}$) assuming HI 21cm absorption by the neutral IGM (from
\cite{carilli02}.
Three different redshifts are displayed: 
$z=12$ (red-light), $z=10$ (violet-medium), and $z=8$ (blue-dark). 
The abscissa for the upper panel is velocity, while that for the
middle and lower panels is the corresponding comoving physical scale. 
The middle panel shows the kinetic ({\it bold lines\/}) and the spin
({\it thin lines\/}) temperature of the 
the neutral IGM over the range of distances that contribute to the
velocity range indicated on the top panel.
The bottom panel shows
the neutral hydrogen density.}
\end{figure}

Most of the variations in transmissivity seen in Figure 1 are due to
mild density inhomogeneities with typical values of the cosmic
overdensity $\delta\sim10$, precisely the structures that at later
times give rise to the Ly$\alpha$ forest.  Because these structures
are typically filamentary, they are at first shock-heated to about 100
K. At $z<10$ resonant Ly$\alpha$ scattering further increases the gas
temperature to several hundred degrees. At the same time the first HII
regions start to appear - one of them manifests itself in the sharply
lower neutral hydrogen density at the right edge of the bottom panel
at $z=8$.

The optical depth, $\tau$, of the neutral hydrogen to 21cm absorption
is: 
\begin{equation}\label{tau}
  \tau = 0.008 \left(T_{\rm CMB}\over T_S\right)
  \left(1+z\over 10\right)^{1/2}x_{\rm HI}(1+\delta),
  \label{tauofz}
\end{equation}
where $x_{\rm HI}$ is the neutral hydrogen fraction, and $\delta$ is
the cosmic overdensity \cite{tozzi00}.  Figure 2 shows in two
panels the joint distribution of the spin temperature, $T_S$, and
neutral hydrogen density (in units of the mean hydrogen density),
$x_{\rm HI}(1+\delta)$, of gas along random lines of sight at two
redshifts: $z=12$ and $z=8$. These parameters, plus $T_{\rm CMB}$,
dictate the HI 21m optical depth via equation (2).

\begin{figure}
\includegraphics[width=2.5in]{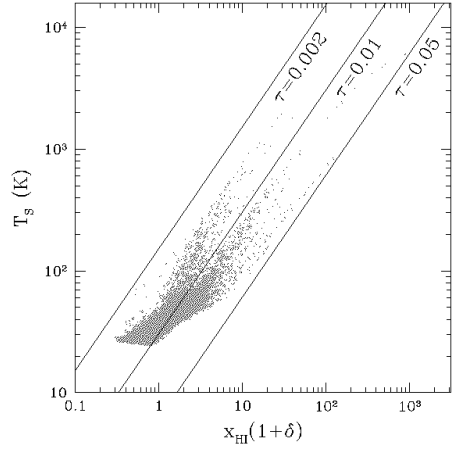}
\includegraphics[width=2.5in]{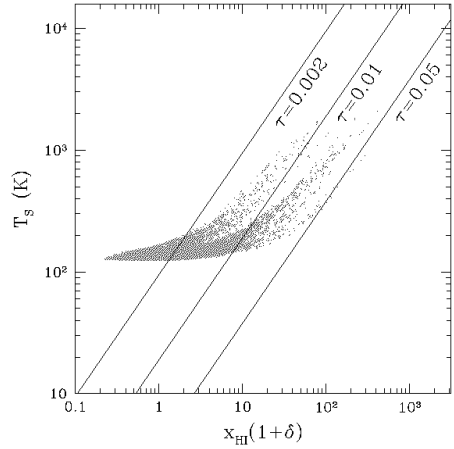}
\caption{The joint distribution of the spin
temperature, T$_{\rm S}$, and neutral hydrogen density, $x_{\rm HI}$(1
+ $\delta$) (in units of the mean hydrogen density) of gas along
random lines of sight at two redshifts: ({\it a\/}) $z=12$ (upper
panel) and ({\it b\/}) $z=8$ (lower panel).  The solid curves indicate
constant HI 21cm optical depth, $\tau$, as labeled \cite{carilli02}.}
\end{figure}

The solid lines in Figure 2 are iso-$\tau$ curves, for $\tau = 0.002$,
0.01, and 0.05.  At $z = 12$ most of the IGM has $\tau \sim 1\%$, and
some of the high density regions reach $\tau \sim 5\%$ and above (a
finite size of the computational box limits our ability to trace the
high $\tau$ tail of the distribution).  These higher $\tau$ points
typically have relatively narrow velocity widths, as low as a few km
s$^{-1}$ (Figure 1), implying HI column densities of order 10$^{19}$
to 10$^{20}$ cm$^{-2}$.  By $z \sim 8$, Ly$\alpha$ heating of the low
density gas increases the mean spin temperature to above 100 K and the
mean IGM optical depth has dropped to $\tau \sim 0.1\%$ (although
cf. \cite{chen03}). However, narrow, higher $\tau$ absorption lines
that form in the still neutral filaments are still easily
identifiable.
 
\subsection{Simulated spectra of high redshift radio sources}

These cosmological simulations can be used to generate synthetic HI
21cm absorption spectra, including the noise characteristic of the
SKA.  For the SKA parameters we assume an effective area of
$5\times10^5$ m$^2$ at 200 MHz, two orthogonal polarizations, and a
system temperature of 250 K (100 K from the receiver and 150 K from
diffuse Galactic emission\footnote{The contribution to system
temperature from diffuse Galactic nonthermal emission behaves as
frequency$^{-2.75}$ in this frequency range.}).  We also make the
simplifying assumption that the ratio of effective area to system
temperature remains roughly constant down to 100 MHz, as could arise
in the case of a low frequency array composed of dipole antennas, and
hence that the sensitivity is constant across the frequency range of
interest (100 MHz to 200 MHz).  We adopt a long, but not unreasonable,
integration time of 10 days (240 hours). These parameters lead to an
expected rms noise level of 34$\mu$Jy in a 1 kHz spectral channel.  We
assume that the correlator will provide at least 10$^4$ spectral
channels over a 10 MHz band, implying a channel width of 1 kHz = 2 km
s$^{-1}$.  We also assume that the spectral bandpass determination
will be at least as good as current telescopes, and presumably
considerably better (10$^4$ over 10's of MHz).

The sources being considered correspond to powerful radio galaxies for
which the emission mechanism is non-thermal (synchrotron) radiation
from a relativistic plasma of electrons and magnetic fields. In most
such sources the spectrum can be described well by a power-law over
the frequency range of interest (0.6 GHz to 3 GHz in the rest frame).

Figure 3a shows a simulated spectrum at 1 kHz resolution of a $z = 10$
radio source with a flux density of 20 mJy at an observing frequency
of 120 MHz. The implied luminosity density at a rest frame
frequency of 151 MHz is then $P_{151} = 2.5\times10^{35}$ erg s$^{-1}$
Hz$^{-1}$.  The on-set of HI 21cm absorption by the neutral IGM is
clearly seen at 129 MHz. The general continuum level drops by about
1$\%$ at this frequency due to the diffuse neutral IGM.  Deeper narrow
lines are also visible to frequencies as high as 170 MHz. At around
130 MHz ($z = 9.9$) there are roughly 5 narrow lines with $\tau \ge
0.02$ per unit MHz, while at 160 MHz ($z=8.9$) the redshift-density
is lower by a factor of 10 or so.

\begin{figure}
\includegraphics[width=2.5in]{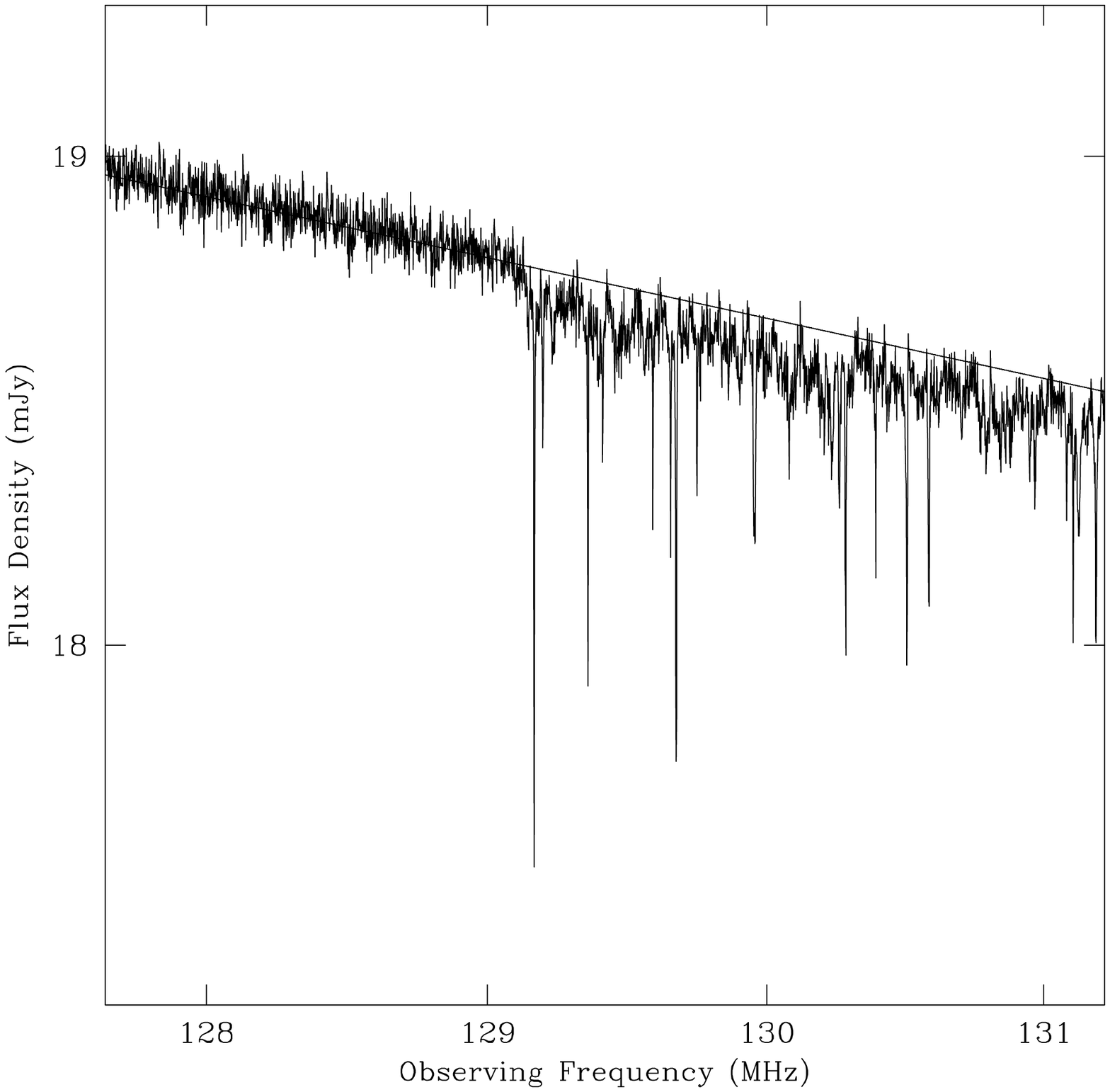}
\includegraphics[width=2.5in]{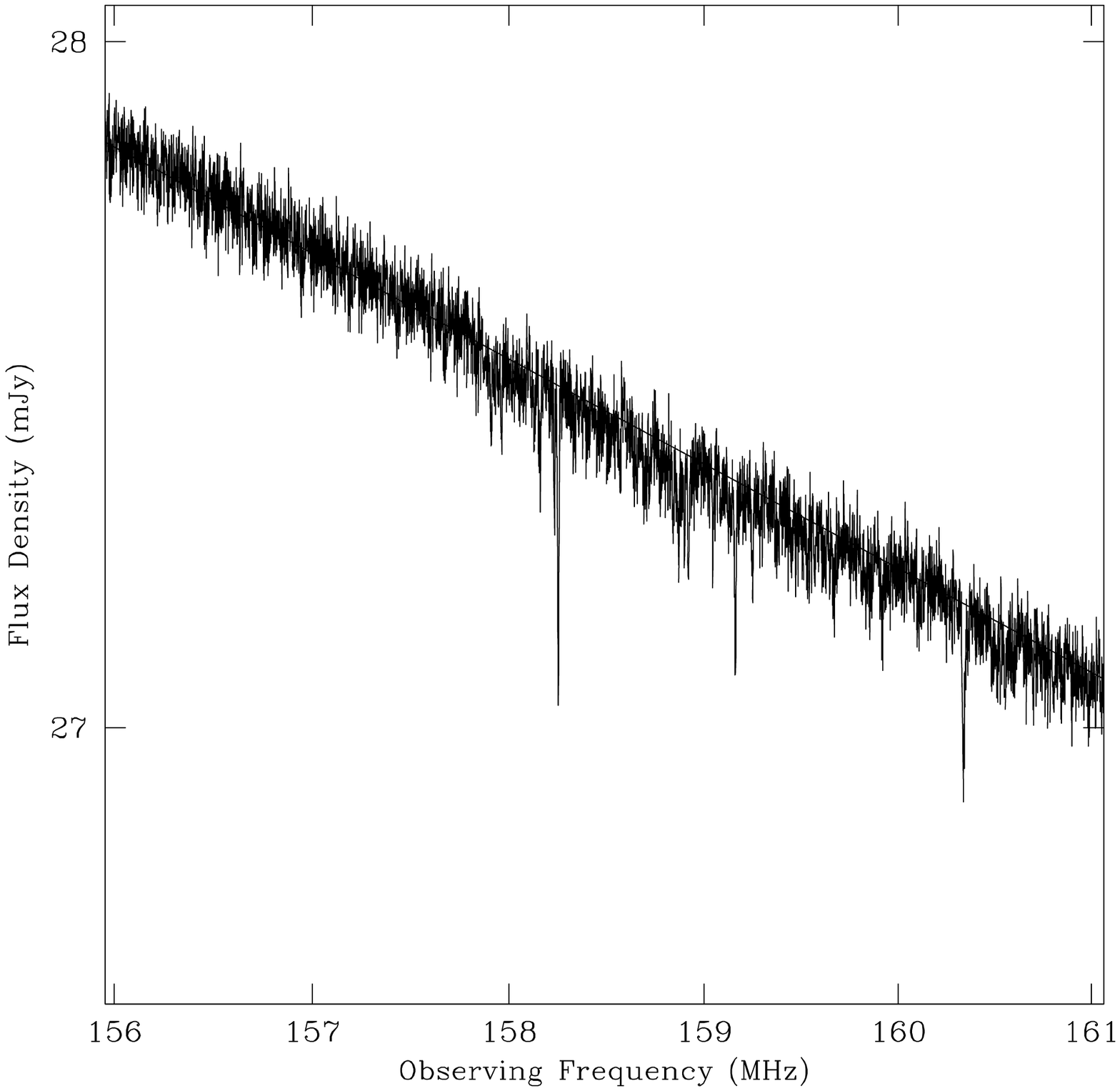}
\caption{{\bf upper panel:} The simulated spectrum 
of a source with S$_{120}$ = 20 mJy at $z = 10$ using the
a spectral model based on Cygnus A 
and assuming HI 21cm absorption by the IGM \cite{carilli02}.
Thermal noise has been added using the specifications of
the SKA and assuming 10 days integration with 1 kHz wide spectral
channels.  The onset of absorption by the neutral IGM is seen
at 129 MHz, corresponding to the HI 21cm line at $z=10$. 
{\bf lower panel} --- Same as 3a, but for a
source with S$_{120} = 35$ mJy at $z = 8$.}
\end{figure}

Figure 3b shows a simulated spectrum at 1 kHz resolution of a $z = 8$
radio source with S$_{120}$ = 35 mJy, again corresponding to $P_{151}
= 2.5\times10^{35}$ erg s$^{-1}$ Hz$^{-1}$. The depression in the
continuum due to absorption by the diffuse IGM is much less evident
than at higher redshift, with a mean value of $\tau \sim 0.1$\%. The
deep narrow lines are still easily seen, but at lower redshift-density
than is found at higher redshifts.

Limits to the detection of absorption by the neutral IGM are
considered in detail in \cite{carilli02}.  In terms of detecting the
deeper narrow lines, the line redshift-density is such that we expect
about two to three narrow lines with $\tau \ge 0.02$ in the range $z =
7$ to 8. In order to detect such lines at 5$\sigma$ at the sensitivity
levels considered herein requires a continuum source of S$_{120}$ = 11
mJy.  The most effective method for detecting absorption by the
neutral IGM is to look for a change in the rms noise level in the
spectrum as a function of frequency, since a sharp increase of the rms
noise level above the system noise is expected at the onset of HI 21cm
absorption.  These simulations suggest we should be able to detect the
on-set of HI 21cm absorption by the IGM at the 5$\sigma$ level toward
a source with S$_{120}$ = 6.5 mJy at $z = 8$.

\subsection{Physical diagnostics}

It is important to emphasize that the structures giving rise to HI
21cm absorption prior to the epoch of reionization are qualitatively
different from those seen after the universe reionizes. After
reionization the HI 21cm lines arise only in rare density peaks
($\delta > 100$) corresponding to (proto)galaxies, ie. the damped Ly
$\alpha$ systems.  Prior to the epoch of fast reionization the
bulk of the IGM is neutral with a measurable opacity in the HI 21cm
line.  The absorption seen in Figure 3 arises in the ubiquitous
`cosmic web', as delineated after reionization by the Ly $\alpha$
forest \cite{bond96}.  The point is simply that the
Ly $\alpha$ forest as seen after the epoch of reionization corresponds
to structures with neutral hydrogen column densities of order
$10^{13}$ cm$^{-2}$ to $10^{15}$ cm$^{-2}$, and neutral fractions of
order $10^{-6}$ to 10$^{-4}$ \cite{weinberg97}.  Before
reionization these same structures will then have neutral hydrogen
column densities of order $10^{19}$ cm$^{-2}$ to $10^{20}$ cm$^{-2}$,
and hence may be detectable in HI 21cm absorption.

In general, the spectra shown in Figure 3 show that the structure of
the neutral IGM during the EoR is rich in temperature, density,
and velocity structure. Studying this structure via HI 21cm absorption
lines will offer important clues to the evolution of the IGM at the
very onset of galaxy formation. In particular, the recent discovery of
polarization of the CMB on large angular scale \cite{kogut03}
suggests that reionization may be a complex process, extending from $z
\sim 20$ down to $z \sim 6$.  Such complex reionization could alter
substantially the thermal state of the neutral IGM, thereby changing
the absorption properties. For instance, Cen \cite{cen03} concludes
that HI 21cm absorption may be an ideal diagnostic to test for an
early epoch of reionization.

\section{Absorption by collapsed structures}

\subsection{Mini-halos}

In addition to the filaments and sheets that are part of the cosmic
web, 21 cm absorption offers the opportunity to observe individual
collapsed structures.  The most important of these are so-called
`minihalos,' which are collapsed halos whose virial temperatures are
too small for atomic hydrogen cooling to be efficient ($T_{\rm vir} <
10^4$ K).  Without cooling, the gas is unable to form stars, and it
collapses only to a characteristic overdensity $\sim 200$.  These
minihalos therefore become small absorbing clouds in the IGM,
analogous to HI clouds in our own galaxy.\footnote{If molecular
hydrogen cooling is efficient, some minihalos can cool to form stars.
The efficiency of H$_2$ formation in the early universe is
controversial \cite{loeb01}, and is
likely only to be effective at extremely high redshifts ($z>20$), when
minihalos are still rare.}

Furlanetto \& Loeb \cite{furlanetto03} have recently considered 21 cm
absorption from these objects (see Fig. 4).  A typical minihalo
absorption line has an optical depth of a few percent, assuming that
the line of sight passes within the virial radius of the halo, and a
width of a few km/s (determined by the virial temperature of the
minihalo).  They used a simple structure formation model to predict
the number density of minihalo absorption lines, $dN/dz \sim 4$--$10$,
comparable to the expected number of absorbers from the cosmic web.
(Note that minihalos are too small to be resolved by most numerical
simulations, including those discussed above, so they constitute an
independent absorption mechanism.)  Distinguishing the two kinds of
absorbers (mini-halos vs. cosmic web) may be difficult, although
clustering and line profiles may help.

Limits on the abundance of minihalo absorbers would place stringent
constraints on structure formation models.  Furlanetto \& Loeb showed
that the abundance of minihalo absorption lines is essentially
determined by two parameters.  First, their number density is
sensitive to the temperature of the IGM, which determines the Jeans
mass.  In the absence of IGM heating, simple structure formation
models predict that more than $10\%$ of the gas should sit in
minihalos by $z \sim 12$.  As the IGM is heated, however, the Jeans
mass approaches the cooling threshold and minihalos become
significantly more rare, for reasons similar to those that cause the
declining number density of absorption features from the cosmic web.
Cen \cite{cen03} and Oh \& Haiman \cite{oh03} have argued that
minihalos are therefore good indicators of `fossil HII regions,' or
regions that have been ionized and subsequently recombined: because
the temperature and entropy remain high, such regions would lack
minihalo absorbers.

The second parameter is the extragalactic Ly$\alpha$ background, which
affects $T_S$ in the outskirts of minihalos (see \S 2.1; the inner
regions are generally dense enough for collisional coupling between
the spin and kinetic temperaturs of the gas to be efficient).  The UV
background therefore determines the number density of weak absorbing
features (with $\tau < 1\%$), as shown in Fig. 4.  The three panels
show the same minihalo field as the extragalactic Ly$\alpha$
background flux is dialed up.  The weak features rapidly disappear.
Thus, like the cosmic web, minihalo absorption is also
a sensitive probe of the radiation background through its effect on
$T_S$.

\begin{figure}
\includegraphics[width=3in]{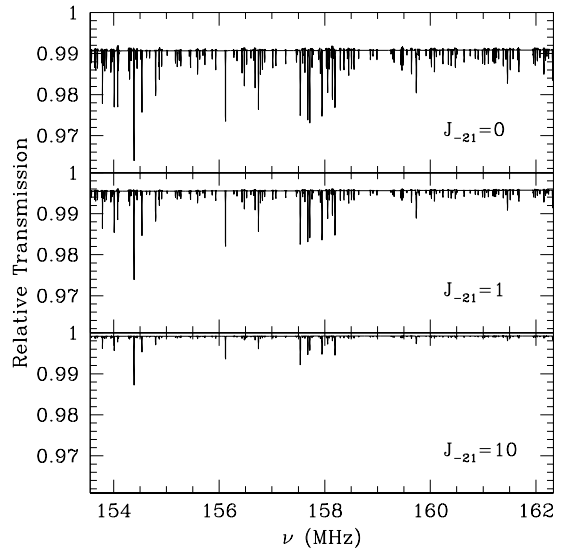}
\caption{Simulated spectra for HI 21cm absorption by mini-halos at $z
= 9$. The different frames show the expected absorption for different
values of the extragalactic Ly$\alpha$ radiation field $J_{-21}$, in
units of $10^{-21}$ erg s$^{-1}$ cm$^{-2}$ Hz$^{-1}$ sr$^{-1}$ 
\cite{furlanetto03}.}
\end{figure}

\subsection{Protogalaxies}

Halos with $T_{\rm vir} > 10^4$ K can cool through atomic transitions.
These objects can thus form stars and become ``protogalaxies'' that
are in many ways similar to nearby galaxies.  In particular, they host
dense neutral gas clouds. If a line of sight intersects such a cloud
we would expect strong 21 cm absorption.  Furlanetto \& Loeb
\cite{furlanetto03} also considered these objects.  They made the
simple assumption that these objects form neutral disks in order to
estimate the number density of absorption features.  Not surprisingly,
the lines are rare (with $dN/dz < 0.1$) but quite deep ($\tau > 1$)
and wide ($> 10$ km/s).  Because they are such strong features, one
could use fainter sources in order to search for protogalaxies.
Exceptionally luminous starburst galaxies are one possibility;
gamma-ray bursts with mJy radio afterglows are another.  The latter
are particularly interesting because one would expect to see
absorption from the host galaxy with a high probability, which would
provide invaluable information about the ISM characteristics of
high-redshift star-forming galaxies.

\section{Radio source populations}

The critical requirement for absorption studies is the existence of
radio sources of sufficient surface brightness at very high redshift.
A simple argument in favor of very high-$z$ radio sources is that a
radio galaxy and two radio loud quasars with luminosities comparable
to, or significantly larger than, the values required by the SKA have
already been found between $z = 5$ and 6 
\cite{vanbreugal99,petric03}.  Extrapolating to $z > 6$ seems 
a relatively small step in 
cosmic time. Also, Carilli et al. \cite{carilli02} show that the
physical changes at high $z$ (such as a higher IGM density or high
T$_{CMB}$) do not preclude radio galaxies at these high redshifts.

The question of radio-loud AGN within the EoR has been considered in
detail by Carilli et al. \cite{carilli02}, Haiman et al.
\cite{haiman04}, and Jarvis \& Rawlings (this volume). Carilli et al.,
and Jarvis \& Rawlings, take an empirical approach, based on observed
high $z$ source populations.  The evolution of the luminosity function
for powerful radio sources is reasonably well quantified for sources
with $P_{151} \ge 6\times10^{35}$ erg s$^{-1}$ Hz$^{-1}$ out to $z
\sim 4$ \cite{jarvis01}.  Beyond this redshift existing surveys are
consistent with either a flat comoving number density, or a steep
decline. In either case, a simple extrapolation of the Jarvis et
al. luminosity function to higher redshift and lower luminosity leads
to a substantial number of radio sources beyond the EoR, even in the
very pessimistic case of a space density decreasing exponentially with
redshift. For a flat co-moving number density evolution there will be
many sources ($1.4\times 10^5$ sources between $z = 6$ and 15) in the
sky at high enough redshift such that HI 21cm absorption during the
EoR could be observed. Even in the case of a steeply declining source
population there will still be a reasonable number of sources (2240
sources between $z = 6$ and 15) toward which absorption experiments
can be performed.

\begin{figure}
\includegraphics[width=3in]{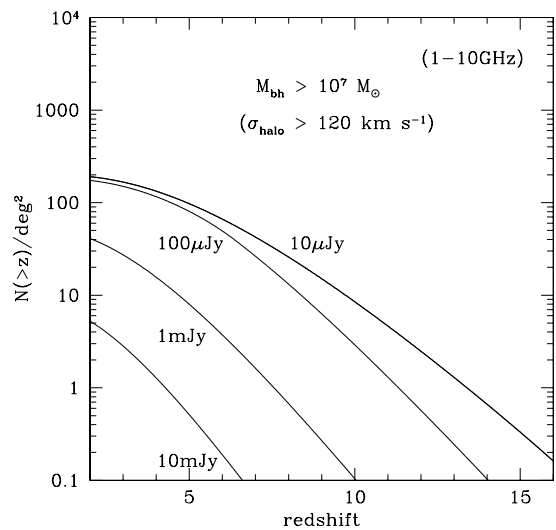}
\caption{Predicted number of radio--loud quasars as a function of
redshift with flux densities greater than the indicated levels,
ignoring the contribution of black holes with masses below $10^7~{\rm
M_\odot}$ \cite{haiman04}. }
\end{figure}

Haiman et al. \cite{haiman04} take a more theoretical approach, based
on the expected evolution of super massive black holes. Again, making
reasonable extrapolations of known AGN populations, and using well
constrained models of structure formation, they predict a significant
number of radio sources into the EoR.  Their predicted areal density of 
high $z$ radio sources is shown in Figure 5. 

Overall, current models and observations of radio-loud AGN evolution
suggest between 0.05 and 1 radio sources per square degree at $z > 6$
with $\rm S_{150MHz} \ge 6$ mJy, adequate for EoR HI 21cm absorption
studies with the SKA (see Jarvis \& Rawlings, this volume, for
details).

\section{An SKA search for HI 21cm absorption by the neutral IGM}

Even if luminous radio sources exist during the EoR, will we be
able to identify them?  An obvious method would to color-select
radio-loud sources in the near-IR, eg.  I-band drop-outs at $z = 7$ or
H-band dropouts at $z = 15$.  This could be done using near-IR
observations of mJy radio samples or radio observations of near-IR
samples.  However, this method requires the objects be bright at
(rest-frame) UV to blue wavelengths, thereby limiting the sample to
dust-poor, optically luminous sources.  

A potentially more fruitful method is to use the radio data
itself. For instance, Carilli et al. \cite{carilli02} show that a
running-rms test could be used to detect sources with anomalously
large noise values in the relevant frequency range of 100 MHz to 200
MHz due to the onset of 21cm absorption by the neutral IGM. The
initial identification could be made by comparing the rms for a given
source to the typical rms derived from all the field sources. Once a
potential high-$z$ candidate is identified, then a second test could
be done to see how the noise behaves as a function of frequency in the
candidate source spectrum.  In this way one might also derive the
source redshift from the on-set of HI 21cm absorption by the neutral
IGM.

What survey area is required, and how many sources need to be
considered in order to find a radio source in the EoR?  For the most
optimistic model (flat comoving number density evolution), the
analysis of section 4 showed that there should be about 3 sources
deg$^{-2}$ at $z > 6.5$ with sufficient radio flux density to detect
HI 21cm absorption by the IGM using the SKA.  The more pessimistic
redshift cut-off model based on luminous QSO evolution leads to 0.05
sources deg$^{-2}$.  The counts of celestial sources with S$_{1400}
\ge 1$ mJy have been determined by a number of groups, and all values
are consistent with: $\rm N(>S_{1400}) = (0.010\pm0.002)~
S_{1400}^{-1.0\pm0.15}$ arcmin$^{-2}$, with the 1400 MHz flux density,
S$_{1400}$, in mJy \cite{white97}.  At the
relevant flux density limits (S$_{1400} \sim 0.5$ mJy, assuming
$\alpha^{1400}_{150} = -1$), the surface
density of all celestial radio sources is then about 72 sources
deg$^{-2}$.  The implied ratio of sources beyond the epoch of
reionization to foreground sources is then about $1\over{25}$ in the
flat evolution model and $1\over{1400}$ in the QSO cut-off model. In
either case it appears to be a tractable sifting problem.

Another important unknown parameter in any 21 cm forest search is the
ionization history itself.  Both \cite{carilli02} and 
\cite{furlanetto03} showed that the absorption signatures decrease in
number density and strength as one approaches complete reionization.
This occurs primarily because the IGM temperature is expected to
increase to $T > 300$ K well before reionization, through X-ray and
shock-heating.  This decreases the typical optical depth (through
$T_S$) and, by raising the Jeans mass, decreases the amount of
collapsed and collapsing structures.  Thus we would ideally like
bright radio sources to exist not just at high redshift but a
substantial time before reionization.

Recently, Kogut et al. \cite{kogut03} measured a high optical depth to
electron scattering for the cosmic microwave background, indicating
that reionization may have begun at $z \sim 20$.  If so, it will be
much more difficult to find bright sources before
reionization.  Even if one constrains reionization to end at $z \sim
6$ through substantial recombination or long phases of partial
ionization \cite{cen03,wyithe03,haiman03},
heating should accompany the early phases of reionization.  Typical
lines of sight would resemble Figure 3b more than Figure 3a, even at
$z>10$.  In this case, the \emph{lack} of absorption features in an
otherwise neutral medium (perhaps detected through its 21 cm emission)
would constitute a strong argument for an early era of ionization 
\cite{cen03,oh03}.

\section{Confusion}

Foregrounds have been shown to be a potential problem when studying HI
21cm emission from the neutral IGM, including confusion by continuum
sources in the field \cite{dimatteo02}, and by free-free emission from
the first ionized structures \cite{oh03b}, although a carefully
planned, multifrequency experiment should be able to surmount such
difficulties (see contribution by Furlanetto, this volume).  Studies
of HI 21cm \emph{absorption} do not suffer from these confusion
problems for two reasons: (i) the expected lines are narrow, as
opposed to the very broad (continuum) confusion signal, and (ii)
absorption is best done at high (arcsecond) spatial resolution.

One remaining potential source of confusion in the absorption
experiments are radio recombination lines (RRLs) arising in the first
ionized structures \cite{oh03b}. However, these can be
differentiated from HI 21cm lines by the easily recognizable frequency
dependent structure of RRLs.  Moreover, it is likely that at 1.4 GHz
(rest frame) the RRLs may appear as stimulated emission, rather than
absorption. Indeed, RRL absorption, or stimulated emission, seen
toward high $z$ radio loud sources may constitute another interesting
probe of the ionized IGM during the EoR.

\end{document}